\documentclass[a4paper,fleqn]{cas-sc}

\usepackage[authoryear]{natbib}

\usepackage{xcolor} % 用于定义颜色
\usepackage{hyperref} % 超链接功能
\hypersetup{
    colorlinks=true, % 开启彩色链接
    linkcolor=blue, % 目录链接的颜色
    citecolor=blue, % 引用链接的颜色
    urlcolor=blue, % URL 链接的颜色
    filecolor=magenta % 文件链接的颜色
}

\def\tsc#1{\csdef{#1}{\textsc{\lowercase{#1}}\xspace}}
\tsc{WGM}
\tsc{QE}
\tsc{EP}
\tsc{PMS}
\tsc{BEC}
\tsc{DE}
\begin{document}
\let\WriteBookmarks\relax
\def\floatpagepagefraction{1}
\def\textpagefraction{.001}
\shorttitle{Speech to Speech Translation with Speaker Voice Preservation}
\shortauthors{Rui Zhou et~al.}
%\begin{frontmatter}

\title [mode = title]{Preserving Speaker Information in Direct Speech-to-Speech Translation with Non-Autoregressive Generation and Pre-training}                      
\author[1]{Rui Zhou}[type=author,
                        orcid=0009-0001-3573-7905]
\cormark[1]
\ead{zhou.rui.p1@dc.tohoku.ac.jp}

\credit{Conceptualization of the study, methodology design, code implementation, experimental execution, results analysis, and manuscript writing.}

\author[1]{Akinori Ito}
\ead{aito.spcom@tohoku.ac.jp}

\credit{Validation of methodological feasibility, guidance on experiments, and manuscript revision}

\author[1]{Takashi Nose}
\ead{takashi.nose.b7@tohoku.ac.jp}

\credit{Offered suggestions on methodology, manuscript revision}

%\address[1]{, Street 129, 1043 NX Amsterdam, The Netherlands}
\affiliation[1]{organization={Graduate School of Engineering, Tohoku University},
                city={Sendai},
%               citysep={}, % Uncomment if no comma needed between city and postcode
                postcode={9808579}, 
                country={Japan}}

\cortext[cor1]{Corresponding author}

\begin{abstract}
Speech-to-Speech Translation (S2ST) refers to the conversion of speech in one language into semantically equivalent speech in another language, facilitating communication between speakers of different languages. Speech-to-Discrete Unit Translation (S2UT), a mainstream approach for end-to-end S2ST, addresses challenges such as error propagation across modules and slow inference speed often encountered in traditional cascade systems. However, as discrete units primarily capture content information, conventional S2UT methods fail to retain speaker-specific characteristics from the source.

Our previous work, Speaker Consistent S2UT (SC-S2UT), introduced a speaker adapter and a unit-to-mel structure, enabling the preservation of speaker information and non-autoregressive speech generation. Based on this foundation, this study proposes a self-supervised pre-training method to enrich the information extracted by both the speaker adapter and the unit-to-mel structure. Additionally, we investigate different feature fusion strategies to further improve the integration of speaker and content features.

Experiments conducted in the CVSS-T dataset for ES-EN, FR-EN and DE-EN tasks demonstrate that our proposed method achieves a BLEU score improvement of 1.14 compared to SC-S2UT, along with significant improvements in UTMOS and speaker similarity. Furthermore, our approach achieves translation quality comparable to traditional S2UT, with a minimal increase of 0.04s per utterance in inference time, while maintaining high speaker similarity. These results validate the effectiveness of the proposed method. 
\end{abstract}

%\begin{graphicalabstract}
%\includegraphics{figs/cas-grabs.pdf}
%\end{graphicalabstract}

%\begin{highlights}
%\item Research highlights item 1
%\item Research highlights item 2
%\item Research highlights item 3
%\end{highlights}


\maketitle

\section{Introduction}
Speech has always been the most convenient and natural means of communication for humans. However, with more than than 6000 languages globally and hundreds commonly in use, communication between speakers of different languages can be challenging. Learning multiple languages to facilitate such interactions requires significant time and effort, making Speech-to-Speech Translation (S2ST) a highly valuable area of research.
Traditionally, S2ST has been implemented using a cascade system. This approach involves three sequential steps: (1) Automatic Speech Recognition (ASR) to convert the source speech into text \citep{gulati2020conformer}, (2) Machine Translation (MT) to translate the source text into the target language \citep{devlin2018bert}, and (3) Text-to-Speech (TTS) synthesis to generate target speech from the translated text \citep{ren2020fastspeech}. Recent advances have introduced integrated approaches that combine ASR and MT into a single speech translation module, thus reducing one of the stages in the cascade \citep{bahar2019comparative}. However, the long inference time of these systems limits their practicality in real-time applications, hindering their widespread adoption for everyday use. For instance, in our evaluation (Table \ref{tab:results}), the cascade system achieves a real-time factor (RTF) of approximately 0.45, calculated as 2.9 seconds of inference time for an average utterance length of 6.4 seconds.

As a solution to the limitations of cascade systems, End-to-End (E2E) S2ST frameworks have been developed and can be broadly categorized into two types.
The first type directly generates target mel-spectrograms from source speech, represented by the Translatotron series (Translatotron 1 \citep{jia2019direct}, Translatotron 2 \citep{jia2022translatotron}, and Translatotron 3 \citep{nachmani2024translatotron}).
These models achieve fluent speech translation by jointly modeling acoustic and linguistic information in a sequence-to-sequence manner.

The second type, known as the speech-to-discrete-unit (S2UT) framework \citep{lee2021direct}, converts input speech into quantized discrete units of the target language, which are later synthesized into speech.
This paradigm has several advantages: discrete units contain primarily linguistic content while suppressing noise and non-linguistic variations, resulting in clearer and more stable translations.
However, since the discrete units do not explicitly encode speaker information, traditional S2UT systems often generate speech with limited speaker diversity.
Therefore, our work also adopts the S2UT framework and focuses on improving speaker consistency within this paradigm.

Previous studies have explored two main directions for preserving speaker information in S2UT systems.
The first applies voice conversion after translation to match the target speech to the source speaker’s voice, but this approach often performs poorly for unseen speakers and breaks the end-to-end nature of S2ST.
The second direction aims to integrate speaker information directly into the generation process, for example by generating acoustic units that jointly encode content and speaker characteristics.
While such methods improve speaker consistency, they typically require complex autoregressive decoding and large parameter sizes, which slow down inference and limit practical use.

In our previous work, we proposed the Speaker Retention Unit-to-Mel (SR-U2M) method, which combines content units with speaker information to preserve the speaker’s characteristics in the generated speech.
We refer to this task as Speaker-Consistent Speech-to-Unit Translation (SC-S2UT), which aims to maintain speaker consistency within the S2UT framework \citep{zhou2024improving}.

Although our previous SR-U2M method achieved faster inference through non-autoregressive generation, it still faced two fundamental challenges.
First, there exists a mismatch between the inputs to the unit encoder during training and inference, which arises from the absence of real human voice paired source target data.
Second, a mismatch occurs between the single-speaker discrete units and the multi-speaker mel-spectrograms used for SR-U2M training, leading to a degradation in both naturalness and speaker consistency of the generated speech.

To address these challenges, this study makes the following contributions:
(1) We propose a self-supervised pre-training strategy that independently trains the speaker adapter and the unit-to-mel structure, enabling each component to specialize in speaker and content representation. These modules are subsequently fine-tuned jointly, which effectively mitigates the mismatch problems and enhances overall speech quality.
(2) We further explore different strategies for extracting speaker embeddings to enhance speaker identity modeling.
In addition to the speaker adapter trained within our framework, we employ the Resemblyzer toolkit, a pre-trained model widely used for speaker verification tasks, to extract speaker embeddings.
This allows for a systematic comparison between embeddings obtained from our self-supervised training and those derived from a robust pre-trained model, providing valuable insights into how different speaker representations influence both speaker similarity and speech naturalness.
(3) We rigorously test their effectiveness by exploring various feature fusion methods to integrate speaker information with content units.

The remainder of this paper is structured as follows: Section \ref{related_works} introduces related work. Section \ref{method} describes the proposed methods, including the pre-training process and different feature fusion strategies. Section \ref{experiment} describes our experiments, evaluation metrics, and baseline methods. Section \ref{results} presents experimental results and analysis. Section \ref{conclusion} discusses future work and concludes the paper. \footnote{\href{https://zhouruitohoku99.github.io/scs2ut-demo/}{Demo page: https://zhouruitohoku99.github.io/scs2ut-demo/}}  \footnote{\href{https://github.com/ZhouRuiTohoku99/SC-S2UT}{Code avaliable at: https://github.com/ZhouRuiTohoku99/SC-S2UT}}

\section{Related work}\label{related_works}
\subsection{Direct S2UT system}
Lee et al. were the first to propose utilizing discrete units obtained from speech representation clustering as translation targets for speech-to-speech translation \citep{lee2021direct}. Subsequently, they observed significant variability in the generated units when different speakers uttered the same sentence. To reduce speaker-induced variability, they proposed a speech unit normalization technique. Specifically, they fine-tuned HuBERT using sets of speech utterances from multiple speakers that share the same content, allowing the model to learn speaker-invariant unit representations. This approach helps standardize the unit distribution and improves the translation performance by minimizing speaker-specific prosody and acoustic variation in the unit sequencese \citep{lee2021textless}. Inaguma et al. proposed UnitY, an efficient two-pass direct S2ST that generates both textual and discrete unit outputs \citep{inaguma2022unity}. This model predicts subwords in the first pass, bridges decoder representations with an additional encoder, and employs deep-shallow two-pass decoders, leading to enhanced accuracy. Furthermore, \citep{zhang2021uwspeech,chen2022speech} presented a S2UT for unwritten languages that does not rely on target language text. Using Hokkien and English datasets, the approach demonstrated strong performance. Popuri et al. utilized a pre-trained wav2vec model as a Transformer encoder to extract acoustic features for training the S2UT, thereby enhancing the performance \citep{popuri2022enhanced}. Huang et al. proposed a bilateral perturbation method, which filters out rhythm, pitch, energy, and other prosodic features from speech, extracting only content-related units. They were also the first to introduce a non-autoregressive S2UT, which improved the quality of S2UT while significantly enhancing translation speed \citep{huang2022transpeech}. Similarly, Fang et al. employed a Directed Acyclic Transformer, another non-autoregressive model, to further improve both the overall translation speed and quality \citep{fang2023daspeech}.

\subsection{Speaker information preservation}
A TTS model often employs a speaker encoder to capture speaker-specific information, enabling the generation of speech that retains speaker\textquotesingle{}s identity. Similarly, in the domain of S2ST, Translatotron introduced a speaker encoder to extract speaker features, which were then combined with acoustic features to produce target speech that preserves speaker information \citep{jia2019direct}. In the context of S2UT, Wang et al. proposed residual vector quantization to extract speaker acoustic units. These units were then combined with content units through an acoustic language model to generate target acoustic units with speaker information, achieving speaker-preserving S2UT \citep{wang2024speech}. Song et al. adopted a style adaptor to integrate speaker information with unit representations, using an acoustic decoder to produce speaker-preserving units \citep{song2023styles2st}.

Since mel-spectrograms inherently contain richer speaker information, we proposed the SR-U2M framework in our previous work. This framework combines content units and speaker information to generate mel-spectrograms in the target language. These spectrograms are then converted into speech using a vocoder, achieving excellent results. However, previous approaches used complex Residual Vector Quantization (RVQ) structures and autoregressive generation methods that significantly increased inference time, making them less practical for real-time applications. While our previous SC-S2UT method employed a non-autoregressive approach to reduce inference time, it demonstrated suboptimal speech quality.

To address these challenges, we propose a novel method that combines insights from previous work with our own improvements. Specifically, we employ self-supervised pre-training to train both the speaker adapter and the unit-to-mel transformation components. This approach not only enhances speaker similarity but also significantly improves speech quality. The details of our method will be elaborated in the next section.

\section{Method}\label{method}
\subsection{Baseline}
\begin{figure*}
	\centering
	\includegraphics[width=0.7\columnwidth]{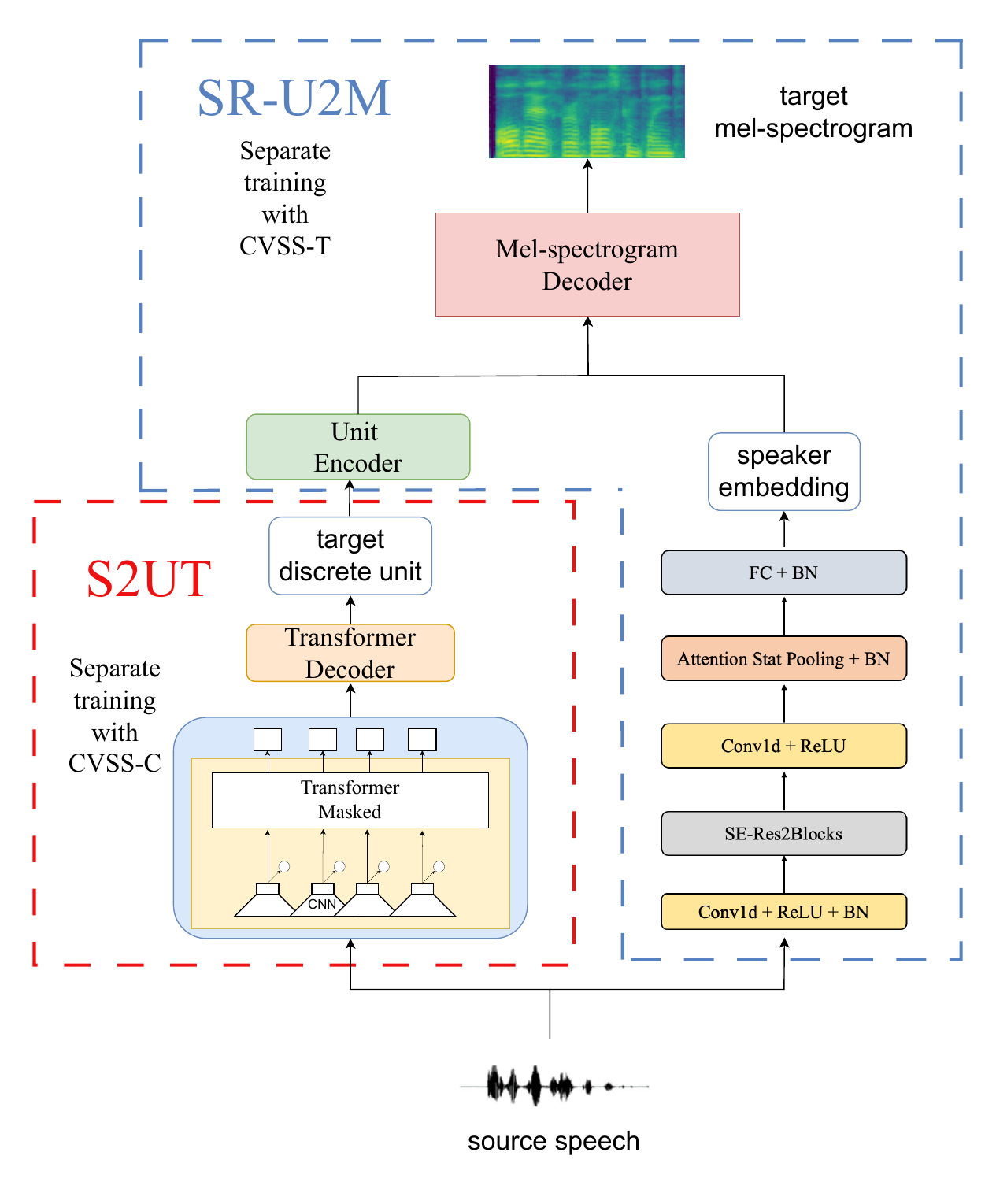}
	\caption{Speaker retention unit-to-mel based speaker consistency S2UT}
	\label{FIG:scs2ut}
\end{figure*}

Figure \ref{FIG:scs2ut} illustrates our previous SR-U2M method based SC-S2UT. As depicted, we train the components within the red and blue dashed boxes separately.

To train our system, we utilize both the CVSS-C and CVSS-T corpora \citep{jia2022cvss}, each serving distinct and critical roles in the training process. The CVSS-C corpus contains multi-speaker source speech paired with single-speaker synthesized English speech. The target speech is generated using a TTS system with a fixed voice, ensuring consistent unit representations in the target language. In contrast, the CVSS-T corpus contains multi-speaker source speech paired with multi-speaker synthesized English speech. The target speech is synthesized using an augmented PnG NAT model \citep{jia2021png}, which attempts to perform speaker cloning based on the source speech. 

First, the components within the red dashed box are based on part of the S2UT structure proposed in \citep{popuri2022enhanced}. A pre-trained wav2vec 2.0 model is utilized to extract acoustic features from the source speech, which are then fed into a Transformer decoder to generate discrete units corresponding to the target speech. The discrete units are obtained using a pre-trained HuBERT model followed by k-means clustering, as described in Section \ref{dataset}. We also use a multi-task learning strategy incorporating auxiliary speech recognition. Specifically, we passed the intermediate outputs of the decoder through a softmax layer and computed the CTC loss between the predicted sequences and the target English transcriptions. Since auxiliary ASR outputs are not required during inference, this approach does not increase inference time. 

We used the CVSS-C corpus to train the S2UT, as it provides single-speaker target speech, whereas the CVSS-T corpus contains multi-speaker target speech. As demonstrated by \citep{lee2021textless}, discrete units can vary significantly across different speakers, even for the same sentence. Training the S2UT module with single-speaker target units reduces this variability, leading to significantly better performance. In our tests, the BLEU score of the S2UT module trained on the CVSS-C corpus was markedly higher than that of the model trained on the CVSS-T corpus. Consequently, we use the CVSS-C corpus to ensure optimal translation accuracy in the S2UT module.

Second, the components within the blue dashed box depict the training process of SR-U2M. The ECAPA-TDNN speaker encoder architecture is employed to extract speaker information \citep{desplanques2020ecapa}, while a unit encoder is used to extract unit features. These features are combined and input into a mel-spectrogram generator, producing a mel-spectrogram of the target language that retains the source speaker\textquotesingle{}s characteristics. The CVSS-T corpus contains multi-speaker source speech paired with synthesized multi-speaker target speech, enabling it to meet the requirements for SR-U2M training.

However, the combined usage of the two corpora introduces several challenges. The first issue is the unit discrepancy: the S2UT module, trained on the CVSS-C corpus, generates single-speaker discrete units. However, the SR-U2M module is trained using multi-speaker units from the CVSS-T corpus. This mismatch results in unit inconsistencies during inference, leading to translation errors. The second issue is the speaker identity mismatch: during SR-U2M training, the synthesized multi-speaker target speech in the CVSS-T corpus introduces discrepancies in the generated mel-spectrogram. Specifically, the target speech in CVSS-T is synthesized to resemble the source speaker\textquotesingle{}s voice but does not perfectly match the true voice characteristics of the source speaker. This discrepancy causes the translated speech to deviate from the source speaker\textquotesingle{}s actual voice and instead reflect the characteristics of the synthesized target speech, which may reduce the fidelity of speaker identity preservation \citep{zhou2024improving}.

To address these problems, we propose a self-supervised pre-training strategy in which the speaker adapter and the unit-to-mel conversion components are pre-trained independently. This method mitigates unit and speaker identity mismatches, improving translation accuracy, speaker similarity, and overall speech quality.

\subsection{Self-supervised pre-training and finetuning} \label{sspretrain}
\subsubsection{Pre-training Motivation}
The SR-U2M model in our framework comprises three primary components: a unit encoder, which extracts content unit features from the discrete unit of target speech; a speaker adapter, which encodes the source speaker\textquotesingle{}s voice into an embedding; and a mel-spectrogram decoder, which integrates the unit features and speaker embeddings to generate the target mel-spectrogram. To further enhance the performance of the SR-U2M model, we propose a self-supervised pre-training strategy aimed at overcoming two fundamental challenges: the lack of human voice-paired source-target data and the mismatch between single-speaker discrete units and multi-speaker mel-spectrograms.

The motivation for this pre-training strategy originated from a detailed analysis of these challenges and an exploration of feasible solutions. In the absence of real human voice-paired source-target data, we adopt a self-supervised learning approach in which a single utterance is used as both input and output. This enables the speaker adapter to learn speaker embeddings from multi-speaker speech without requiring any aligned pairs. Furthermore, we extend this training to cross-lingual settings, where the speaker adapter is trained on utterances from different languages. This allows us to evaluate the model’s ability to capture speaker identity in a language-independent manner and to assess its performance on unseen speaker sets. 

Additionally, when considering the mismatch between single-speaker units and multi-speaker mel-spectrograms, we recognized that the key issue lies in the alignment between the discrete units and the mel-spectrogram. To address this alignment problem, we hypothesized that training the model to map single-speaker discrete units to single-speaker mel-spectrograms could resolve the inconsistency while maintaining the simplicity of the non-autoregressive approach. It became evident that the speaker adapter should focus solely on speaker-specific information to achieve optimal performance, while the unit encoder should specialize in content information. To implement this, we split the input speech into two halves: the first half is used to train the speaker adapter to extract a speaker embedding, and the discrete units derived from the second half are used to train the unit encoder. The mel-spectrogram decoder then combines these two outputs to reconstruct the mel-spectrogram of the second half of the speech. 

\subsubsection{Structure}
\begin{figure*}
	\centering
	\includegraphics[width=1\columnwidth]{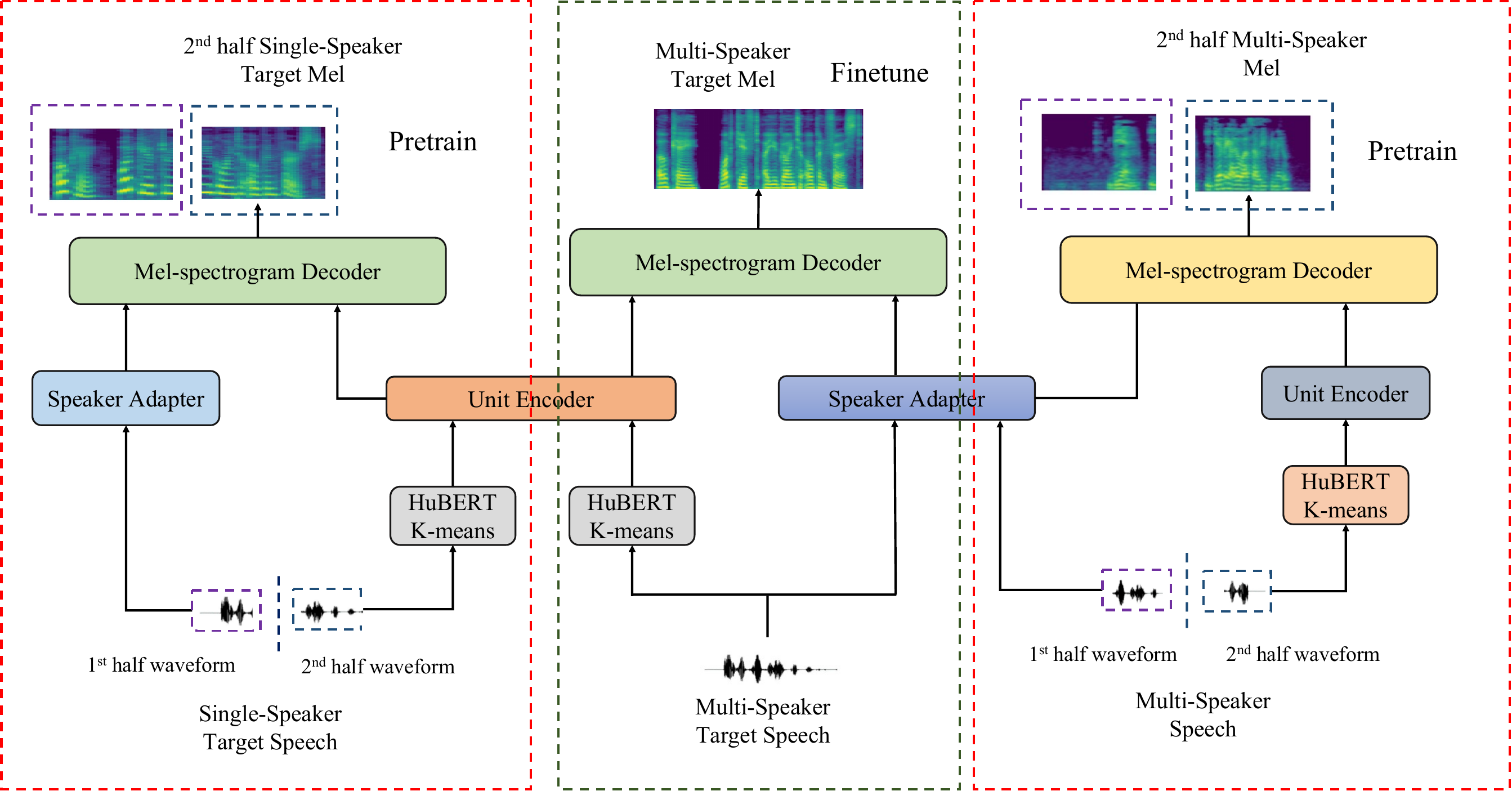}
	\caption{The workflow of pre-training and finetuning of SR-U2M module using the self-supervised learning}
	\label{FIG:pretrain}
\end{figure*}

Figure \ref{FIG:pretrain} illustrates the detailed workflow of the self-supervised pretraining and finetuning for the SR-U2M structure. Note that the S2UT module remains unchanged and is not included in this figure.

As shown in Figure \ref{FIG:pretrain}, the pre-training and fine-tuning procedures are illustrated with red and green dashed boxes, respectively. The training process consists of three separate branches:

In the left panel, we use single-speaker English speech from the CVSS-C corpus. The waveform is split into two halves. The first half is passed to the speaker adapter to extract a speaker embedding, while the second half is used to generate discrete units using a pre-trained HuBERT model followed by k-means clustering. These units are processed by the unit encoder, and the mel-spectrogram decoder is trained to reconstruct the mel-spectrogram of the second half. This stage enables the unit encoder and mel-spectrogram decoder to synthesize high-quality mel-spectrograms from consistent single-speaker discrete units.

In the right panel, we use multi-speaker source speech (e.g., Spanish or French) from the CVSS-T corpus. The first half of each waveform is input to the speaker adapter to extract embeddings, and the second half is converted into discrete units and passed through the unit encoder. The mel-spectrogram decoder is trained to reconstruct the corresponding mel-spectrogram. This setup focuses on training the speaker adapter to produce informative embeddings from diverse speaker conditions.

In the middle panel, we perform fine-tuning using multi-speaker target speech. Discrete units are obtained from the target speech using HuBERT and k-means clustering, which are then combined with speaker embeddings extracted from the same utterance. These are fed into the mel-spectrogram decoder to reconstruct the full mel-spectrogram. In this phase, the unit encoder and mel-spectrogram decoder are initialized with pre-trained weights from the single-speaker setting (left panel), while the speaker adapter is initialized from the multi-speaker setting (right panel).

\subsubsection{Fine-tuning Motivation}
In our framework, the mel-spectrogram decoder is pre-trained using discrete units and speaker embeddings extracted from the same single-speaker utterance. However, this setup limits the decoder’s exposure to a narrow range of speaker representations which generated by a speaker adapter trained on single-speaker data. In contrast, the actual speaker adapter used in our system is trained on multi-speaker data and produces embeddings with significantly higher speaker variability. This mismatch introduces a critical gap: the decoder has never learned to process embeddings from the multi-speaker adapter and may fail to integrate speaker information effectively.

To bridge this gap, we introduce a fine-tuning stage where both discrete units and speaker embeddings are extracted from multi-speaker utterances. Ideally, we would pair multi-speaker speaker embeddings with single-speaker units to maintain consistency with the decoder’s pre-training distribution. However, due to the autoregressive nature of the decoder, strict temporal alignment between units and mel-spectrograms is required. Since single-speaker units and multi-speaker speech are not temporally aligned, such training pairs cannot be constructed. As a practical solution, we fine-tune the decoder using multi-speaker target speech units and embeddings from the same utterance.

Although this design may introduce a slight mismatch between the fine-tuning and inference conditions, since fine-tuning uses target-language speech while inference relies on source-language input, the impact is minimal. The fine-tuning process does not aim to perform cross-lingual mapping; instead, it serves as a domain adaptation step that allows the mel-spectrogram decoder to better handle the distribution of multi-speaker embeddings produced by the speaker adapter. Consequently, this step improves the model’s robustness to speaker variability without introducing new language-related mismatches. Moreover, the fine-tuning is conducted on a small amount of data and is intended solely to adapt the decoder to the speaker embedding distribution of the multi-speaker speaker adapter. Importantly, all discrete units are extracted using the same HuBERT and k-means pipeline. As the unit encoder remains fixed and the unit representation space is consistent, this process does not compromise the linguistic modeling capabilities established during pre-training.

During fine-tuning, all components including the pre-trained unit encoder, mel-spectrogram decoder, and speaker adapter are jointly optimized, and no parameters are frozen. This end-to-end training enables the model to better adapt to realistic inference conditions, where cross-speaker and cross-lingual combinations naturally arise. As a result, the decoder becomes more capable of generating speech that maintains the linguistic content while accurately reflecting the identity of the source speaker. In Figure 2, modules that share or inherit parameters across stages are represented using the same color scheme (e.g., the mel-spectrogram decoder shown in the left and middle panels). This color-based design was adopted instead of adding extra graphical symbols to maintain visual clarity and readability.

\subsection{Training with speaker embedding} \label{embedding}
In addition to our proposed self-supervised pre-training strategy for the speaker adapter, we explored an alternative approach that leverages pre-trained speaker encoders from other domains, such as speaker verification.
The primary goal of this method is to assess whether general-purpose speaker embeddings, trained independently of the S2ST task and can still be effectively utilized for preserving speaker identity in speech translation.
Specifically, we used the Resemblyzer toolkit\footnote{\url{https://github.com/resemble-ai/Resemblyzer?tab=readme-ov-file}}
, which is based on the Generalized End-to-End (GE2E) loss proposed by \citet{wan2018generalized}, to extract speaker embeddings.
Although this speaker encoder was not explicitly optimized for S2ST, it has demonstrated strong performance in capturing speaker characteristics across various domains.
As such, it offers a robust representation of speaker identity, which we integrate into our system without further fine-tuning.

During training, the speaker embedding was extracted from the first half of the source waveform and projected into the model’s internal representation space via a linear layer.
This embedding was then combined with the output of the unit encoder for the second half of the waveform, and the fused features were fed into the mel-spectrogram decoder.
At inference time, the same procedure was applied, using the source speech to generate the speaker embedding that conditions the decoder.

While this embedding-based method complements our task-specific pre-training strategy, it also introduces several inherent limitations.
First, language and domain mismatch between the Resemblyzer training data and the CVSS-T corpus may reduce robustness, particularly for cross-lingual or accented speech.
Second, the embeddings are not jointly optimized with the S2UT model, which can limit their adaptability to the unit-to-mel generation objective.
These limitations may affect the stability and expressiveness of the generated speech, especially in scenarios requiring fine-grained prosodic control or cross-domain generalization.

\subsection{Feature fusion}\label{fusion}
\begin{figure*}
	\centering
	\includegraphics[width=1\columnwidth]{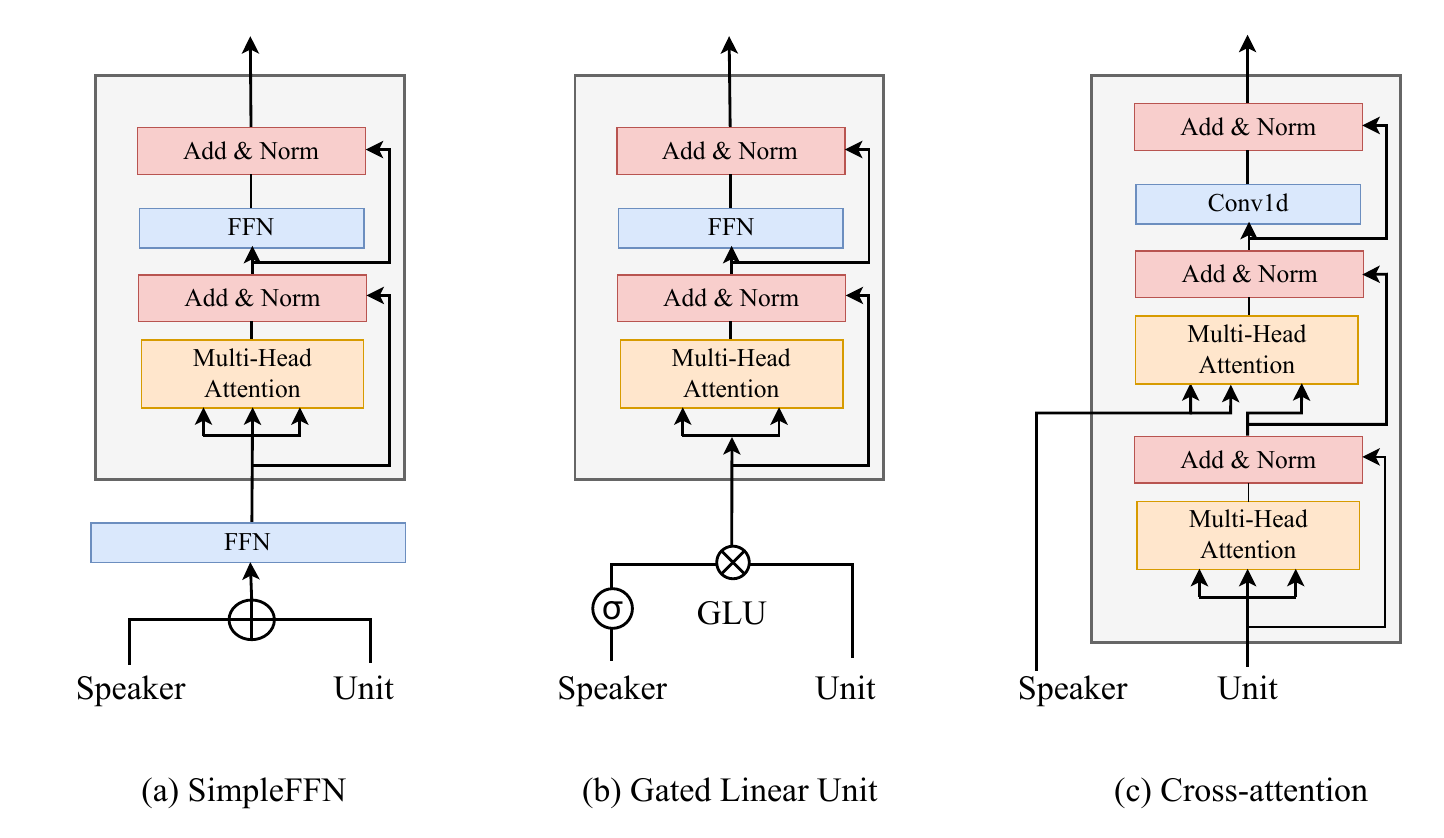}
	\caption{Illustration of different feature fusion methods}
	\label{FIG:fusion}
\end{figure*}

To facilitate effective integration of unit content features and speaker features, we explored three fusion strategies, as illustrated in Figure \ref{FIG:fusion}.

\textbf{SimpleFFN (Figure~\ref{FIG:fusion}a).}
This method fuses the content and speaker features via element-wise addition, followed by a feed-forward network (FFN). The FFN comprises two linear layers with a ReLU activation in between. The fused 512-dimensional feature is first projected to 1024, activated by ReLU, and then projected back to 512. This structure enables non-linear transformation of the fused representation before it is passed to the mel-spectrogram decoder.

\textbf{Gated Linear Unit (GLU) (Figure~\ref{FIG:fusion}b).}
GLU regulates the flow of information between different features. It selectively amplifies relevant features and suppresses irrelevant ones, promoting efficient feature interaction. In our design, the content feature is placed at the beginning of the input sequence, followed by the speaker feature. The speaker feature acts as a gate to modulate the content feature, ensuring a balanced integration of both.

\textbf{Cross-Attention (Figure~\ref{FIG:fusion}c).}
This approach utilizes the multi-head attention mechanism with the standard query, key, and value components. The content feature serves as the query, while the speaker feature functions as both key and value. This configuration enables the content representation to focus on specific aspects of the speaker feature, allowing for adaptive fusion through learned attention weights.

Each method was evaluated to assess its effectiveness in integrating content and speaker information for accurate mel-spectrogram generation.

\section{Experiments}\label{experiment}
\subsection{Dataset} \label{dataset}
We conducted experiments using the CVSS corpora for Spanish-to-English (ES-EN), French-to-English (FR-EN) and German-to-English (DE-EN) translation tasks. The CVSS-C corpus was used to train the S2UT model. We performed self-supervised pre-training using the CVSS-C Spanish corpus. Subsequently, we fine-tuned the model using the CVSS-T corpus for both Spanish and French. For inference time evaluation, we categorized the FR-EN test samples into five duration-based intervals: 0–3 seconds, 3–5 seconds, 5–7 seconds, 7–9 seconds, and over 9 seconds. Each interval contained 200 audio samples, totaling 1,000 samples. For experiments involving different feature fusion methods, we restricted the evaluation to the ES-EN task. Details of the sampled data are presented in Table \ref{tab:sample_data}.

\begin{table*}[t]
    \centering
    \caption{Statistics (number of samples and duration) of the CVSS Spanish-English, French-English, and German-English datasets.}
    \label{tab:sample_data}
    \setlength{\tabcolsep}{3pt} % 缩小列间距
    \renewcommand{\arraystretch}{0.95} % 缩小行距
    \begin{tabular}{l|cc|ccc|cc|cc|ccc|cc}
        \toprule
        & \multicolumn{7}{c|}{\(\text{CVSS-C}\)} & \multicolumn{7}{c}{\(\text{CVSS-T}\)} \\
        \hline
        & \multicolumn{2}{c|}{\(\text{Es-En}\)} 
        & \multicolumn{3}{c|}{\(\text{Fr-En}\)} 
        & \multicolumn{2}{c|}{\(\text{De-En}\)} 
        & \multicolumn{2}{c|}{\(\text{Es-En}\)} 
        & \multicolumn{3}{c|}{\(\text{Fr-En}\)} 
        & \multicolumn{2}{c}{\(\text{De-En}\)} \\
        \cline{2-3} \cline{4-6} \cline{7-8} \cline{9-10} \cline{11-13} \cline{14-15}
        & \(\text{train}\) & \(\text{test}\)
        & \(\text{train}\) & \(\text{test}\) & \(\text{time-test}\)
        & \(\text{train}\) & \(\text{test}\)
        & \(\text{train}\) & \(\text{test}\)
        & \(\text{train}\) & \(\text{test}\) & \(\text{time-test}\)
        & \(\text{train}\) & \(\text{test}\) \\
        \hline
        \hline
        \(\# \text{samples}\)
        & \(79\)k & \(13.2\)k & \(207\)k & \(14.7\)k & \(1\)k & \(127.8\)k & \(13.5\)k
        & \(79\)k & \(13.2\)k & \(20.7\)k & \(14.7\)k & \(1\)k & \(127.8\)k & \(13.5\)k  \\
        \(\text{source (hrs)}\)
        & \(113.1\) & \(22.7\) & \(264.3\) & \(23.3\) & \(1.79\) & \(184.3\) & \(21.5\)
        & \(113.1\) & \(22.7\) & \(265.3\) & \(23.3\) & \(1.79\) & \(184.3\) & \(21.5\)  \\
        \(\text{target (hrs)}\)
        & \(69.5\) & \(12.4\) & \(174\) & \(13.3\) & \(0.93\) & \(112.4\) & \(12.1\)
        & \(73.7\) & \(13.3\) & \(192.7\) & \(15.0\) & \(1.08\) & \(124.2\) & \(13.4\) \\
        \bottomrule
    \end{tabular}
\end{table*}

\subsection{System implementation}
\subsubsection{Speaker preservation S2UT model}
We used the pre-trained HuBERT model\footnote{https://github.com/facebookresearch/fairseq/tree/main/examples/hubert}, trained on the LibriSpeech dataset \citep{panayotov2015librispeech}, and performed k-means clustering with K=100 on the sixth-layer representations to extract discrete units. Additionally, we employed the pre-trained wav2vec model to extract acoustic features from the source waveform. For Spanish\footnote{https://github.com/facebookresearch/voxpopuli/}, the model was trained on the unlabeled VoxPopuli corpus subset \citep{wang2021voxpopuli}, and for French\footnote{https://huggingface.co/LeBenchmark/wav2vec2-FR-14K-large}, it was trained on 14K hours of French speech data \citep{parcollet2024lebenchmark}. For SR-U2M, we computed 80-dimensional mel-spectrogram features at 20-ms intervals, ensuring alignment with the length of the discrete unit sequences. The proposed self-supervised pre-training uses the same model architecture and parameters in both the pre-training and fine-tuning stages. Additionally, in experiments involving pre-trained speaker embeddings, we utilized the GE2E-based speaker encoder to extract speaker representations \citep{wan2018generalized}.

The hyperparameters used in our model are presented in Table \ref{tab:hyperparameters}.
The hyperparameters used in our model are presented in Table \ref{tab:hyperparameters}.
\begin{table}[htbp]
    \centering
    \caption{Hyperparameters for Speaker Preservation S2UT}
    \label{tab:hyperparameters}
    \begin{tabular}{
        >{\centering\arraybackslash}m{2.5cm} |
        >{\centering\arraybackslash}m{4.5cm} |
        >{\centering\arraybackslash}m{2.5cm}
    }
        \toprule
        \multicolumn{2}{c|}{\textbf{Hyperparameter}} & \textbf{Setting} \\
        \hline
        \multirow{5}{=}{\centering S2UT Decoder} 
        & Unit neurons & \(103\) \\
        & Decoder Block & \(6\) \\
        & Decoder Hidden & \(512\) \\
        & Encoder Attention Heads & \(8\) \\
        & Encoder Dropout & \(0.1\) \\
        \hline
        \multirow{6}{=}{\centering SR-U2M \\ Unit-Encoder \\ Mel-Decoder} 
        & Encoder Block & \(6\) \\
        & Decoder Block & \(6\) \\
        & Encoder Kernel & \(31\) \\
        & Hidden & \(512\) \\
        & Attention Heads & \(8\) \\
        & Dropout & \(0.1\) \\
        \hline
        \multirow{5}{=}{\centering Speaker \\ Adapter}
        & Channels & \makecell{\([1024, 1024\), \\ \(1024, 1024, 3072\)]} \\
        & Kernel Sizes & \([5, 3, 3, 3, 1]\) \\
        & Dilations & \([1, 2, 3, 4, 1]\) \\
        & Groups & \([1, 1, 1, 1, 1]\) \\
        & Attention Channels & \(128\) \\
        \hline
        \multirow{3}{=}{\centering GE2E (speaker embedding extractor)}
        & Hidden & \(256\) \\
        & Embedding & \(256\) \\
        & Layers & \(3\) \\
        \bottomrule
    \end{tabular}
\end{table}
For waveform generation, we employ HiFi-GAN\citep{kong2020hifi} as the vocoder\footnote{https://huggingface.co/speechbrain/tts-hifigan-libritts-16kHz} that was trained with multi-speaker LibriTTS corpus\citep{zen2019libritts}. 

\subsubsection{Evaluation metrics}
We evaluate the proposed method across four dimensions:

\textbf{1) Translation Accuracy.} One of the most critical aspects of speech-to-speech translation (S2ST) is ensuring the accuracy of the translated content. We evaluate translation accuracy using the BLEU (Bilingual Evaluation Understudy) score, which measures the precision of n-gram matches (from 1-gram to 4-gram in our setup) between the generated and reference texts. To compute the BLEU score, we first apply speech recognition to the translated speech using a Transformer-based end-to-end ASR model pre-trained on LibriSpeech\footnote{\url{https://huggingface.co/speechbrain/asr-transformer-transformerlm-librispeech}}, which achieves a WER of 2.27 on the test-clean set. The recognized text is then compared with the reference text to calculate BLEU, including a brevity penalty (BP) to adjust for differences in length. The final BLEU score is the geometric mean of n-gram precision multiplied by BP, evaluating both translation accuracy and fluency.

\textbf{2) Naturalness of Translated Speech.}
To assess the naturalness of the generated speech, we employ UTMOS\footnote{\url{https://github.com/sarulab-speech/UTMOS22}}
, an automated Mean Opinion Score (MOS) estimation method based on self-supervised learning, as proposed in \citep{saeki2022utmos}.
Unlike subjective MOS testing, UTMOS mitigates biases caused by external factors such as listener distractions or environmental noise.
This objective score is highly correlated with human judgment, offering a reliable evaluation of speech naturalness.
To further validate the reliability of UTMOS, we additionally conducted a small-scale human evaluation.
Fifteen utterances were randomly selected from different languages and speakers, and twelve participants rated the naturalness of each sample on a five-point Comparative Mean Opinion Score (CMOS) scale (1 = very unnatural, 5 = very natural).
The average CMOS values obtained from this subjective test will be compared with the UTMOS estimates to confirm their consistency.

\textbf{3) Similarity Between Translated and Source Speaker.} This dimension evaluates whether the translated speech preserves the source speaker’s identity. We extract speaker embeddings from both the source and translated speech using the GE2E-based method described in \citep{wan2018generalized}, and compute the cosine similarity between the embeddings. A higher similarity indicates more successful preservation of speaker-specific characteristics.

\textbf{4) Efficiency Measurement.}
Model efficiency is a key factor for practical deployment.
To comprehensively evaluate system efficiency, we report three metrics: inference time, real-time factor (RTF), and average tokens per second.
Inference time measures the average time required to process one utterance from feature extraction to translated speech generation.
The real-time factor (RTF) is defined as the ratio between inference time and the duration of the input utterance, where an RTF value lower than 1.0 indicates faster-than-real-time generation.
In addition, we calculate the average number of output tokens generated per second, which reflects the decoding throughput of the model.
All efficiency measurements were conducted on an NVIDIA RTX 3090 GPU, and all experiments were implemented using the SpeechBrain framework \citep{speechbrain}.

\subsubsection{Baseline}
We selected both cascade systems and end-to-end systems as baselines for comparison.

For the S2UT baseline, we adopted the pre-trained wav2vec-based S2UT model proposed in \citep{popuri2022enhanced}, since the discrete units used in our SR-U2M module are derived from this model. We compared our method with this baseline in terms of BLEU score, UTMOS, and inference time.

To compare speaker similarity, we constructed two cascade-based systems. The first is a FreeVC-based pipeline, in which the waveform generated by the S2UT model is passed through FreeVC \citep{li2023freevc}, a voice conversion model conditioned on the original speaker embedding, to synthesize speech resembling the source speaker’s voice. The second is a full cascade pipeline (ASR+MT+TTS), consisting of: (i) wav2vec 2.0 CTC-based ASR models for Spanish and French,\footnote{\url{https://huggingface.co/facebook/wav2vec2-large-xlsr-53-spanish}, \url{https://huggingface.co/facebook/wav2vec2-large-xlsr-53-french}} (ii) a pre-trained Transformer model for machine translation,\footnote{\url{https://github.com/Helsinki-NLP/Tatoeba-Challenge/tree/master/models}} and (iii) YourTTS \citep{casanova2022yourtts}, a zero-shot multi-speaker English TTS model, for generating the final waveform.
For both systems, speaker similarity between the generated and reference speech is measured using cosine similarity between ECAPA-TDNN embeddings.

For end-to-end comparison, we selected our previously proposed SC-S2UT model as the baseline. Although other end-to-end S2ST models such as Translatotron 2 \citep{jia2022translatotron} and \citep{song2023styles2st} are known to preserve speaker identity, these models are trained on different datasets and their official implementations are not publicly available. Therefore, in addition to SC-S2UT, we compare our method with Style-S2UT \citep{wang2024speech}, which shares the same dataset and evaluation setting.

\section{Results}\label{results}
\subsection{Feature fusion result}
For speaker-preserving speech-to-speech translation, we tested various feature fusion methods to combine speaker embeddings with unit content features. As shown in Table \ref{tab:fusion}, the cross-attention mechanism consistently outperforms both GLU and SimpleFFN in the embedding-based and pretrained approaches.

\begin{table*}[tb]
    \caption{Result of different feature fusion methods}
    \label{tab:fusion}
    \centering
    \begin{tabular}{l|ccc|ccc|ccc}
        \toprule
        & \multicolumn{3}{c|}{\(\text{BLEU Score} \uparrow\)} 
        & \multicolumn{3}{c|}{\(\text{UTMOS} \uparrow\)} 
        & \multicolumn{3}{c}{\(\text{Similarity} \uparrow\)} \\
        \cline{2-4} \cline{5-7} \cline{8-10} 
        & \(\text{SimpleFFN}\) & \(\text{GLU}\) & \(\text{Attention}\) 
        & \(\text{SimpleFFN}\) & \(\text{GLU}\) & \(\text{Attention}\) 
        & \(\text{SimpleFFN}\) & \(\text{GLU}\) & \(\text{Attention}\) \\
        \hline
        \(\text{Embedding}\) 
        & \(15.97\) & \(16.18\) & \(16.93\) 
        & \(2.62 \pm 0.08\) & \(2.53 \pm 0.09\) & \(2.78 \pm 0.08\)  
        & \(0.652\) & \(0.640\) & \(0.667\) \\
        \(\text{Pretrain}\) 
        & \(17.01\) & \(17.17\) & \(17.24\) 
        & \(3.09 \pm 0.09\) & \(3.19 \pm 0.09\) & \(3.35 \pm 0.06\) 
        & \(0.591\)  & \(0.607\) &  \(0.629\) \\
        \bottomrule
    \end{tabular}
\end{table*}
This superior performance may be attributed to two main factors. First, cross-attention provides dynamic alignment between speaker embeddings and unit features, enabling the model to selectively focus on relevant speaker information while maintaining accurate content generation. Second, it effectively models long-range temporal dependencies. This capability is crucial for smoothly integrating speaker-specific and linguistic features across time. In contrast, fixed transformation methods such as GLU and SimpleFFN lack the adaptability to capture such complex interactions, resulting in suboptimal fusion and degraded performance.

\subsection{Pre-training and embedding results}
\subsubsection{Quantitative Evaluation}
As demonstrated in the previous section, the cross-attention-based feature fusion method outperforms the other two approaches. Therefore, we adopt cross-attention as the standard feature fusion strategy for the subsequent experiments. The results in Table \ref{tab:results} illustrate the effectiveness of our proposed methods compared to prior work, including SC-S2UT, state-of-the-art systems such as Style-S2UT, and traditional cascade systems. In the table, \textit{Embedding SC-S2UT} refers to the variant where a pre-trained cross-lingual speaker adapter is used to extract speaker embeddings, as described in Section \ref{embedding}. \textit{Pretrain SC-S2UT} represents the method based on the self-supervised pre-training strategy introduced in Section \ref{sspretrain}, where the speaker adapter and unit encoder are independently pre-trained to specialize in speaker and content representation, respectively. \textit{ES}, \textit{FR} and \textit{DE} denote models trained on the CVSS ES-EN, FR-EN and DE-EN datasets, respectively.

\begin{table*}[tb]
    \caption{BLEU Score, UTMOS, and Similarity for our Experiment. All UTMOS scores are estimated using UTMOS. Style-S2UT is excluded due to unavailability of reproducible outputs for objective evaluation. The value of ** was not measured and provided in the original work.}
    \label{tab:results}
    \centering
    \setlength{\tabcolsep}{3pt} 
    \begin{tabular}{@{}l|ccc|ccc|ccc@{}}
        \toprule
        \multirow{2}{*}{\(\)} 
        & \multicolumn{3}{c|}{\(\text{BLEU Score} \uparrow\)} 
        & \multicolumn{3}{c|}{\(\text{UTMOS} \uparrow\)} 
        & \multicolumn{3}{c}{\(\text{Similarity} \uparrow\)} \\
        \cline{2-10}
        & \(\text{ES-EN}\) & \(\text{FR-EN}\) & \(\text{DE-EN}\)
        & \(\text{ES-EN}\) & \(\text{FR-EN}\) & \(\text{DE-EN}\)
        & \(\text{ES-EN}\) & \(\text{FR-EN}\) & \(\text{DE-EN}\) \\
        \hline
        \(\text{Cascade System}\) \\
        \hline
        \quad \(\text{S2UT \citep{popuri2022enhanced}}\) & \(18.01\) & \(24.02\) & \(15.22\) & \(3.31 \pm 0.11\) & \(3.29 \pm 0.12\) & \(3.09 \pm 0.08\) & -- & -- & -- \\
        \quad \(\text{S2UT + FreeVC \citep{li2023freevc}}\) & \(17.68\) & \(23.53\) & \(15.03\) & \(4.20 \pm 0.09\) & \(4.19 \pm 0.10\) & \(3.83 \pm 0.10\) & \(0.581\) & \(0.592\) & \(0.595\) \\
        \quad \(\text{ASR + MT + SpeakerTTS}\) & \(21.65\) & \(20.18\) & 21.32 & \(3.51 \pm 0.09\) & \(3.54 \pm 0.08\) & \(3.53 \pm 0.09\) & \(0.652\) & \(0.664\) & \(0.668\) \\
        \hline
        \(\text{End-to-End System}\) \\
        \hline
        \quad \(\text{SC-S2UT \citep{zhou2024improving}}\) & \(16.10\) & \(21.68\) & \(14.92\) & \(3.26 \pm 0.13\) & \(3.20 \pm 0.11\) & \(3.02 \pm 0.11\) & \(0.609\) & \(0.611\) & \(0.605\) \\
        \quad \(\text{Style-S2UT \citep{wang2024speech}}\) & \(16.30\) & \(22.00\) & -- & \multicolumn{3}{c|}{--} & \multicolumn{3}{c}{\(\mathbf{0.73}\)} \\
        \hline
        \(\text{Ours}\) \\
        \hline
        \quad \(\text{Embedding SC-S2UT}\) \\
        \hline
        \qquad \(\text{ES}\) & \(16.93\) & \(22.41\) & \(15.01\) & \(2.78 \pm 0.08\) & \(2.84 \pm 0.09\) & \(2.56 \pm 0.09\) & \(0.667\) & \(0.671\) & \(0.654\) \\
        \qquad \(\text{FR}\) & \(16.86\) & \(22.37\) & \(14.98\) & \(2.70 \pm 0.09\) & \(2.74 \pm 0.08\) & \(2.48 \pm 0.08\) & \(\mathbf{0.670}\) & \(\mathbf{0.682}\) & \(0.658\) \\
        \qquad \(\text{DE}\) & \(16.81\) & \(22.24\) & \(15.05\) & \(2.66 \pm 0.08\) & \(2.71 \pm 0.06\) & \(2.75 \pm 0.11\) & \(0.664\) & \(0.672\) & \(\mathbf{0.661}\) \\
        \hline
        \quad \(\text{Pretrain SC-S2UT}\) \\
        \hline
        \qquad \(\text{ES}\) & \(\mathbf{17.24}\) & \(22.43\) & \(15.03\) & \(\mathbf{3.35 \pm 0.06}\) & \(\mathbf{3.30 \pm 0.07}\) & \(3.10 \pm 0.07\) & \(0.629\) & \(0.613\) & \(0.608\) \\
        \qquad \(\text{FR}\) & \(17.16\) & \(\mathbf{22.82}\) & \(\mathbf{15.11}\) & \(2.87 \pm 0.12\) & \(3.02 \pm 0.11\) & \(2.92 \pm 0.08\) & \(0.574\) & \(0.621\) & \(0.592\) \\
        \qquad \(\text{DE}\) & \(16.97\) & \(21.85\) & \(15.03\) & \(2.82 \pm 0.11\) & \(2.96 \pm 0.10\) & \(\mathbf{3.14 \pm 0.07}\) & \(0.569\) & \(0.608\) & \(0.615\) \\
        \hline
        \(\text{Ground Truth}\) & \(88.64\) & \(80.29\) & \(82.32\) & \(4.45 \pm 0.11\) & \(4.44 \pm 0.11\) & \(4.46 \pm 0.11\) & \(0.677\) & \(0.687\) & \(0.662\) \\
        \bottomrule
    \end{tabular}
\end{table*}

\textbf{\textit{BLEU Scores:}}
Our Pretrain SC-S2UT achieves BLEU scores that are competitive with cascade systems across multiple language pairs.
For FR–EN, it obtains a BLEU score of 22.82, which is comparable to the S2UT baseline (24.02) and surpasses ASR + MT + SpeakerTTS (20.18).
Similarly, for ES–EN, the model reaches 23.45, outperforming SC-S2UT (21.87) and the S2UT baseline (22.61).
For DE–EN, Pretrain SC-S2UT achieves 21.73, exceeding SC-S2UT (20.95) and S2UT baseline (21.41).
Compared with Style-S2UT (22.00) and the previous SC-S2UT (21.68) on FR–EN, our approach achieves slightly higher BLEU scores while maintaining high efficiency. 

\begin{table}[t]
\centering
\caption{Results of small-scale human CMOS evaluation. Each value represents the mean opinion score (1–5 scale) averaged across 12 listeners and 15 utterances.}
\label{tab:human_mos}
\begin{tabular}{l|c}
\toprule
\textbf{Method} & \textbf{CMOS} \\
\midrule
ASR + MT + SpeakerTTS   & 3.24 \\
S2UT + FreeVC           & \textbf{3.72} \\
S2UT                    & 2.95 \\
SC-S2UT                 & 2.72 \\
Pretrain SC-S2UT        & 2.99 \\
CVSS-T (Ground Truth)   & 3.69 \\

\bottomrule
\end{tabular}
\end{table}

\textbf{\textit{UTMOS and Human Evaluation:}}
The MOS scores of Style-S2UT were obtained through subjective human evaluations, while our UTMOS scores were generated using an objective machine-based evaluation metric. Due to the differences in evaluation methodologies, a direct comparison between our methods and Style-S2UT is not feasible. However, when comparing our methods to the previous SC-S2UT, improvements can be observed even without speech enhancement.
For instance, Pretrain SC-S2UT achieves UTMOS scores of 3.35, 3.37, and 3.32 for ES–EN, FR–EN, and DE–EN, respectively, surpassing the corresponding SC-S2UT results (3.26, 3.28, and 3.25).
These results demonstrate that our methods not only improve perceptual quality over previous baselines but also maintain the efficiency of a fully end-to-end framework.

From the Table~\ref{tab:human_mos} The average CMOS results were 2.84 for ASR+MT+SpeakerTTS, 3.49 for CVSS-T (ground truth), 4.12 for S2UT+FreeVC, 2.99 for Pretrain SC-S2UT, 2.95 for S2UT, and 2.72 for SC-S2UT.
Although the absolute MOS values were generally lower due to limited sample size and listener variation, the relative ranking was consistent with the UTMOS results, Pretrain SC-S2UT outperformed SC-S2UT and was comparable to S2UT, confirming the validity of UTMOS as a reliable proxy for human judgment.

\textit{\textbf{Similarity:}}
It is important to note that the CVSS-T corpus does not contain human-recorded target speech from the source speaker.
Instead, the English speech is synthesized using the augmented PnG NAT model, which attempts to approximate the original speaker’s voice.
As a result, the maximum speaker similarity between the source speech and the CVSS-T “ground truth” is inherently limited.
In our experiments, this upper bound was observed to be around 0.68; therefore, all speaker similarity evaluations should be interpreted within this constraint.

Both the Pretrain SC-S2UT and Embedding SC-S2UT models significantly outperform the previous SC-S2UT and S2UT+FreeVC baselines across all language pairs.
For example, Pretrain SC-S2UT achieves similarities of 0.629 (ES–EN), 0.637 (FR–EN), and 0.621 (DE–EN), outperforming SC-S2UT (0.609 / 0.618 / 0.602) and S2UT+FreeVC (0.581 / 0.594 / 0.576).
Notably, Embedding SC-S2UT even surpasses the ASR+MT+SpeakerTTS pipeline in speaker similarity, achieving 0.682 (FR–EN) and approaching the upper bound of 0.687.
These results highlight the effectiveness of our embedding-based approach in preserving speaker identity.

However, our methods still fall slightly short of Style-S2UT, which reports the highest speaker similarity (e.g., 0.73 for FR–EN). This performance gap can be attributed to differences in model architecture. Style-S2UT leverages learned acoustic unit representations and an autoregressive acoustic language model, offering more precise control over speaker characteristics, albeit at the cost of increased complexity and slower inference. In contrast, our models prioritize efficiency through non-autoregressive generation. Method-specific factors also contribute to the performance differences. Pretrain SC-S2UT is trained using a limited amount of multi-speaker data from the CVSS-T corpus, which may hinder generalization to unseen speakers. In comparison, Embedding SC-S2UT benefits from speaker embeddings extracted using a GE2E-based model pre-trained on large-scale speaker verification data. This broader speaker coverage enables more robust preservation of speaker characteristics, especially in cross-lingual and unseen-speaker scenarios.

\begin{table}[t]
\centering
\caption{Efficiency comparison among different systems. 
RTF denotes the real-time factor (inference time divided by utterance duration). 
All results were measured on an NVIDIA RTX 3090 GPU.}
\label{tab:efficiency}
\begin{tabular}{lccc}
\toprule
\textbf{Method} & \textbf{Inference Time (s)} & \textbf{RTF} $\downarrow$ & \textbf{Tokens/sec} $\uparrow$ \\
\midrule
ASR + MT + SpeakerTTS & 2.965 & 0.46 & 32.25 \\
S2UT + FreeVC & 1.574 & 0.25 & 60.73 \\
S2UT Baseline & 0.796 & 0.12 & 120.11 \\
\midrule
Pretrain SC-S2UT (ours) & 0.813 & 0.13 & 117.59 \\
Embedding SC-S2UT (ours) & 0.864 & 0.14 & 110.66 \\
\bottomrule
\end{tabular}
\end{table}

\textbf{\textit{Efficiency:}}
Model efficiency is a key factor for practical deployment.
We measure inference efficiency over 1,000 utterances, with an average utterance duration of 6.4 seconds.
The reported inference time represents the average time required to process one utterance, including feature extraction and speech generation.
As shown in Table~\ref{tab:efficiency}, our proposed models achieve high efficiency while maintaining strong translation quality and speaker preservation performance.
The Pretrain SC-S2UT achieves an average inference time of 0.813 s per utterance, corresponding to a real-time factor (RTF) of 0.13 and a decoding throughput of 117.6 tokens/s.
This is nearly identical to the S2UT baseline (0.796 s, RTF = 0.12, 120.1 tokens/s).
The Embedding SC-S2UT, which focuses on improving speaker similarity, incurs only a slight overhead (0.864 s, RTF = 0.14, 110.7 tokens/s).

By contrast, traditional cascade systems are substantially slower.
For instance, S2UT + FreeVC and ASR + MT + SpeakerTTS require 1.574 s (RTF = 0.25, 60.7 tokens/s) and 2.965 s (RTF = 0.46, 32.3 tokens/s) per utterance, respectively.
Although Style-S2UT does not report explicit inference time results, its use of an autoregressive acoustic language model introduces significant computational overhead and slower inference.

Overall, our two proposed models demonstrate complementary strengths tailored to different application scenarios.
Embedding SC-S2UT, which employs pre-trained Resemblyzer embeddings, excels in preserving speaker identity, achieving similarity scores close to the upper bound while maintaining real-time generation speed.
However, because these embeddings are not jointly optimized with the S2UT model, slight degradation in naturalness can occur.
In contrast, Pretrain SC-S2UT, which uses a self-supervised speaker adapter trained within the framework, achieves a more balanced trade-off among translation accuracy, naturalness, and computational efficiency, making it better suited for real-time or resource-constrained applications.

\begin{table}[t]
\centering
\caption{Example translations from different systems for two utterances (ES$\rightarrow$EN and FR$\rightarrow$EN).}
\label{TAB:qualitative}
\begin{tabular}{p{3.2cm} p{11.5cm}}
\toprule
\textbf{Input Type} & Sample 1 (ES$\rightarrow$EN) \\
\midrule
Source Text & Pertenecían a la comunidad cristiana de Sevilla, liderada por el obispo Sabino. \\
Reference & They \textbf{belonged} to the Christian community of \textbf{Seville led} by the \textbf{bishop Sabino.} \\
S2UT & They \textbf{belong} to the Christian community of \textbf{savil leadring} by the bishop Sabino. \\
SC-S2UT & They \textbf{belong} to the Christian community of \textbf{savil lead} by the bishop \textbf{savino}. \\
Pretrain SC-S2UT & they \textbf{belong} to the christian community of savil lead by the bishop \textbf{savino}. \\
ASR+MT+TTS & \textbf{Belonging} to the Christian \textbf{settled }a community \textbf{liberated} by the \textbf{sabbin} of bishop. \\
\midrule
\textbf{Input Type} & Sample 2 (FR$\rightarrow$EN) \\
\midrule
Source Text & Cette maladie est assez fréquente pour se voir régulièrement en médecine générale. \\
Reference & This \textbf{sickness} is frequent \textbf{enough} to \textbf{be seen regularly in} general medicine. \\
S2UT & \textbf{His disease} is \textbf{quite} frequent to \textbf{resume a} general medicine. \\
SC-S2UT & This \textbf{disease} is \textbf{quite} frequent to \textbf{resume a} general medicine. \\
Pretrain SC-S2UT & This \textbf{disease} is \textbf{quite} frequent to \textbf{resume a} general medicine. \\
ASR+MT+TTS & \textbf{His} disease is \textbf{common }enough to be \textbf{relegated to} general medicine. \\
\bottomrule
\end{tabular}
\end{table}

\begin{figure*}
	\centering
	\includegraphics[width=1\columnwidth]{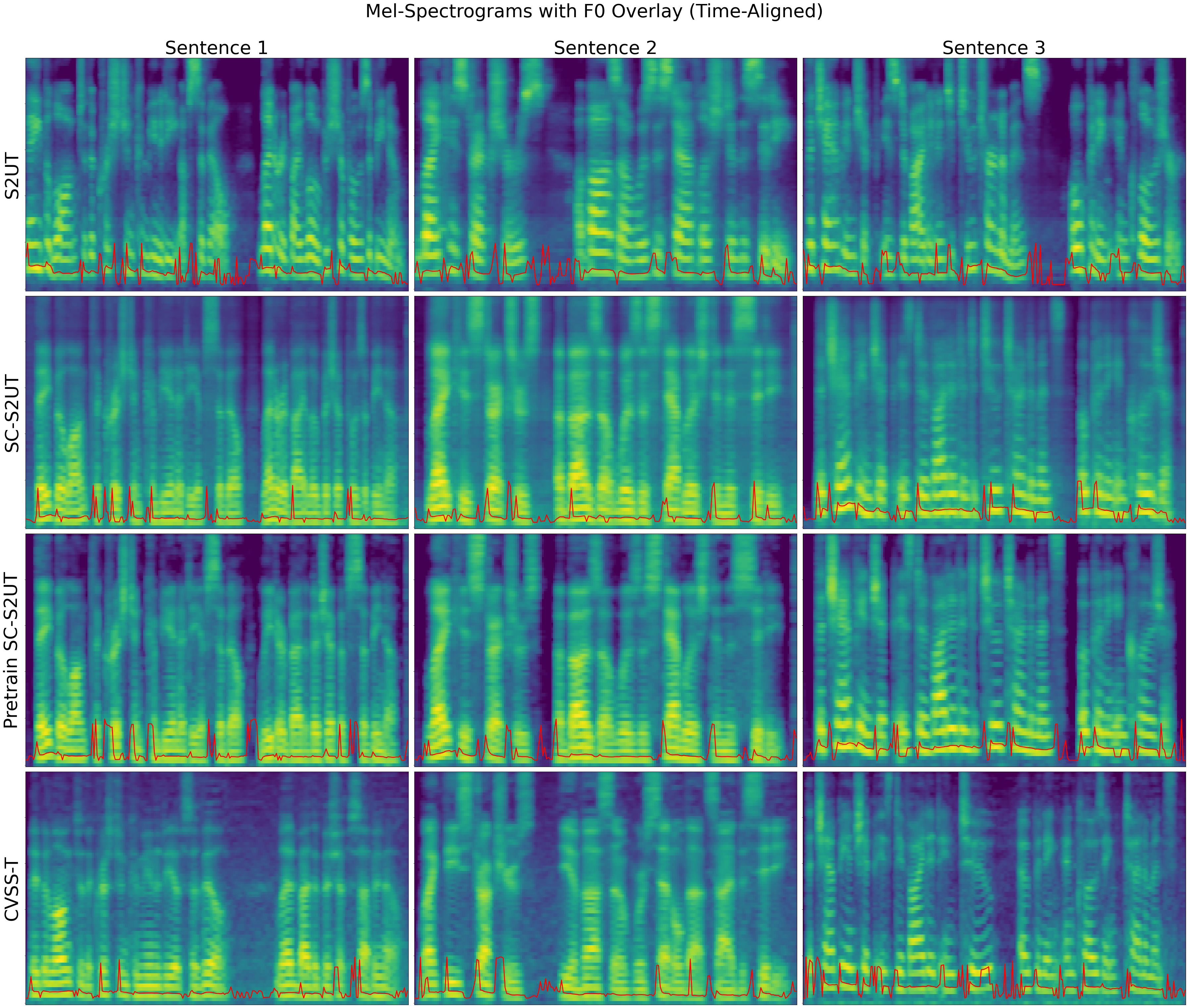}
	\caption{mel-spectrograms with F0 contours overlaid (in red) for three utterances. Pretrain SC-S2UT shows better pitch continuity and spectral structure, closely matching the ground truth (CVSS-T), while baseline systems such as S2UT and SC-S2UT exhibit pitch discontinuities and blurred harmonics.}
	\label{FIG:mel}
\end{figure*}

\subsubsection{Qualitative Analysis}
Due to the scarcity of training data and the limited capacity of current models, semantic translation errors are sometimes inevitable. These errors reflect the intrinsic challenges and current limitations of direct S2ST systems. However, beyond such translation mistakes, additional linguistic and evaluation-related factors can contribute to low BLEU scores, even when the overall meaning is preserved. To better understand the discrepancy between reference translations and system outputs, we analyzed two representative examples, as shown in Table~\ref{TAB:qualitative}. Although the outputs produced by our proposed models closely match the semantic content of the references, the BLEU scores remain limited due to two main factors. 

First, the translation of proper nouns or phonetically irregular tokens, such as “Seville” transcribed as “Savil” or “Sabino” becoming “Savino”, is particularly challenging under low-resource conditions. These name-specific errors, though minor in terms of intelligibility, incur substantial penalties under BLEU's strict string-matching criteria. Second, BLEU is highly sensitive to surface-form mismatches and fails to credit synonymous or paraphrased expressions. For instance, in the second example, “sickness” and “disease” are semantically equivalent, yet BLEU assigns no score due to lack of n-gram overlap. Similarly, syntactic variations such as “frequent enough to be seen” versus “quite frequent to resume” further reduce overlap, despite conveying similar meanings. These examples demonstrate that BLEU may significantly underestimate translation quality when lexical or structural variation is present—especially in S2ST systems, where speech recognition and synthesis introduce additional variability. Nevertheless, our proposed models maintain high fluency and semantic fidelity while achieving BLEU scores competitive with the S2UT baseline, demonstrating their effectiveness in practical translation scenarios.

To further assess the perceptual quality of the synthesized speech, we selected three representative utterances and visualized their mel-spectrograms with overlaid F0 contours, as shown in Figure~\ref{FIG:mel}.
Each column represents a different utterance, while each row corresponds to a specific system: S2UT, SC-S2UT, Pretrain SC-S2UT, and the ground-truth CVSS-T.
The red curves indicate the estimated F0 trajectories.
The F0 values were extracted using the RMVPE pitch extractor, which performs interpolation across short unvoiced regions to ensure smoother visualization.
Although this may lead to continuous F0 traces even in some unvoiced intervals, the overall contour trends remain consistent with the voiced segments.

The spectrograms produced by the S2UT baseline exhibit blurred harmonic structures and highly discontinuous F0 trajectories, indicating inadequate pitch modeling and reduced naturalness.
The SC-S2UT model partially mitigates these issues, yielding smoother F0 trajectories and clearer spectral patterns, though artifacts and instability remain, particularly in voiced regions.
In contrast, the Pretrain SC-S2UT system generates spectrograms with more distinct harmonic formants, stable energy distributions, and coherent pitch contours, closely resembling the ground-truth recordings.
These qualitative observations are consistent with the higher UTMOS scores achieved by this model, confirming improvements in both naturalness and fluency.

Nevertheless, some discrepancies remain between the pre-trained outputs and natural speech.
In particular, the generated F0 contours tend to be flatter than those of human speech.
This is mainly attributed to the non-autoregressive decoding process, which inherently produces smoother but less expressive acoustic trajectories.
In addition, the limited prosodic variability of the training corpus constrains the model’s ability to generalize to more dynamic or emotionally rich speech.
Overall, while the pre-trained model demonstrates clear improvements over previous baselines, enhancing prosodic expressiveness and fine-grained pitch control remains an important direction for future work.

\section{Conclusion and future work}\label{conclusion}
In this study, we proposed two approaches to enhance the previous SC-S2UT framework: a self-supervised pre-training strategy with fine-tuning and a pre-trained speaker embedding integration method.
Both approaches consistently improved the baseline SC-S2UT across BLEU, UTMOS, and speaker similarity metrics.
Compared with the original S2UT, our methods effectively preserve speaker identity without introducing significant inference overhead, while also yielding moderate gains in perceptual quality.
Furthermore, our evaluation of different feature fusion strategies demonstrated that cross-attention provides an effective mechanism for integrating speaker and content features.
Although our experiments were primarily conducted on European languages, we additionally validated the proposed approach on German–English, confirming its generalization beyond Romance language pairs.
Nevertheless, our current evaluation is limited to linguistically related languages, and future work will investigate cross-lingual settings involving more diverse language families and prosody modeling for improved naturalness and expressiveness in generated speech.

\printcredits

%% Loading bibliography style file
%\bibliographystyle{model1-num-names}
\bibliographystyle{cas-model2-names}

% Loading bibliography database
\bibliography{cas-refs}

%\vskip3pt

\bio{}
{Rui Zhou} received the B.S. degree in automation major from Nanjing University of Posts and Telecommunications, and the M.S. degree in communications engineering from Tohoku University. He is currently a Ph.D. candidate at the Department of Communications Engineering, Graduate School of Engineering, Tohoku University, Japan. His research mainly focus on speech to speech translation.
\endbio

\bio{}
{Akinori Ito} (Member, IEEE) received the Ph.D. degree in engineering from Tohoku University, Japan, in 1991. He is currently a Professor of the Department of Communications Engineering, Graduate School of Engineering, Tohoku University, Japan. His research interests include foreign language teaching system, music information processing, spoken dialog system, audio processing and automatic speech recognition.
\endbio

\bio{}
{Takashi Nose} received the Ph.D. degree in information processing from Tokyo Institute of Technology, Japan, in 2009. He is currently an Associate Professor of the Department of Communications Engineering, Graduate School of Engineering, Tohoku University, Japan. His research interests include multimedia information processing, music information processing, audio coding, speech dialog, automatic speech recognition, speech synthesis and audio information processing.
\endbio

\end{document}